\documentclass[twocolumn]{aastex631}
\newcommand{\m}[1]{\mathrm{{#1}}}
\newcommand{\ompe}{\omega_\m{pe}}
\newcommand{\textred}[1]{\textcolor[HTML]{000000}{#1}}
\newcommand{\ctext}[1]{\raise0.2ex\hbox{\textcircled{\scriptsize{#1}}}}
\shortauthors{Kamiido Kazuki and Yutaka Ohira}
\usepackage{natbib}
\usepackage{physics}

\begin{document}

\title{Collisionless Shock Driven by a Supersonic Velocity Shear}
\author[0009-0002-9063-8827]{Kazuki Kamiido}
\author[0000-0002-2387-0151]{Yutaka Ohira}
\affiliation{Department of Earth and Planetary Science, The University of Tokyo, \\
7-3-1 Hongo, Bunkyo-ku, Tokyo 113-0033, Japan}
\email{kamiido-kazuki8990@g.ecc.u-tokyo.ac.jp}

\begin{abstract}
The long-term evolution of a relativistic collisionless velocity shear in an unmagnetized electron-positron plasma is investigated using a first-principle particle-in-cell simulation. 
The Alves instability converts the shear kinetic energy into thermal and magnetic field energy. 
The resulting pressures push the plasma, leading to the formation of collisionless shocks. 
The generated collisionless shocks would accelerate high energy particles, which is a possible solution to the injection problem of shear acceleration. 
In addition, the collisionless shocks generate a magnetic field turbulence that is required for the shear acceleration to work.
\end{abstract}

\section{Introduction}
Velocity shears are ubiquitous in Universe and on Earth, from the flow of a river to black hole accretion disks.
How the velocity shear eventually dissipates is a fundamental problem, especially for the collisionless system. 
In astrophysics, where and how particles are accelerated are also fundamental issues to understand the nonthermal emission from high-energy astrophysical objects and the origin of cosmic rays. 
In a velocity shear, particles can be accelerated if they can move back and forth across the shear layer \citep{Berezhko+81, Ostrowski90, Rieger+06, Ohira13}.
The shear acceleration mechanism requires magnetic-field turbulence to scatter the particle. 
Moreover, the particle energy must be sufficiently large so that the mean free path is larger than the thickness of the shear layer.
Otherwise, the particle cannot be accelerated efficiently (injection problem).
However, the generation mechanism of the magnetic turbulence around the shear layer and the injection problem are still long-standing issues. 
The Kelvin--Helmholtz instability (KHI) and magnetic reconnection could be one of the solutions for a highly magnetized system \citep{Sironi+21}.

The KHI is a well-known instability in a hydrodynamic velocity shear. 
However, astrophysical systems are often filled with collisionless plasmas, in which the mean free path of the Coulomb collision is much larger than the system size. 
In the collisionless two-fluid plasma, there are three unstable modes \citep[GI;][]{Miller+16}: the KHI, the Gruzinov instability \citep{Gruzinov08}, and the Alves instability \citep[AI][]{Alves+15}. 
The wavevectors of the KHI and the GI are parallel to the shear velocity, while the wavevector of the AI is perpendicular to it. 
It should be emphasized that although a supersonic shear flow is stable in the hydrodynamical framework \citep{Miura+82}, it is unstable in the kinetic plasma scale.

The instability in the collisionless supersonic velocity shear was first investigated in 2008 \citep{Gruzinov08}.
Although there are some studies about collisionless velocity shears \citep[e.g.,][]{Alves+12, Grismayer+13, Alves+14, Nishikawa+16, Guo+25, Guo+26}, the history of those studies is relatively short compared to the collisionless shock and the magnetic reconnection. 
Therefore, the parameter dependence and long-term evolution have not been studied sufficiently. 
In this study, we investigate the long-term evolution of a relativistic supersonic velocity shear, where the free energy is much larger than the initial thermal energy. 
We use a first-principle particle-in-cell simulation.
Although some previous studies examined the nonlinear evolution of collisionless relativistic velocity shears \citep{Liang+13, Liang+17, Yao+20}, we first show that collisionless shocks are produced by dissipating the supersonic velocity shear. 
This collisionless shock generates magnetic-field turbulence and will accelerate particles later, which are essential for the shear acceleration to work.

\section{Simulation setup}
We performed a two-dimensional particle-in-cell simulation in the $x$-$y$ plane using the open code, \textit{Wuming} \citep{Wuming}, where we modified the initial condition of this code to suit our simulation. 
Since the AI is the most unstable mode in a relativistic collisionless velocity shear \citep{Alves+15}, we chose the simulation plane perpendicular to the shear velocity ($\vb{U}_0 =U_0 \vb{e}_z$, where $\vb{U}_0$ is the four-velocity of the shear flow). 
We set the initial velocity profile as $U_0/c = +10$ in $L_x/4 < x < 3L_x/4$ and $U_0/c = -10$ in $x < L_x/4$ and $x > 3L_x/4$, where $L_x$ and $c$ are the $x$-directional size of the simulation box and the light speed, respectively.
Periodic boundaries are prepared for both $x$ and $y$ directions. We only visualize the left half of the simulation domain ($0< x < L_x/2$) because the right half is almost the mirror image of the left half. 
As a first step, we consider the simplest system in this work: an unmagnetized electron-positron plasma. 
We set $30$ simulation particles per cell for electrons and positrons. 
The simulation size is $L_x / (c/\ompe) \times L_y / (c/\ompe) = 2400 \times 320$, and the cell size is $\Delta x / (c/\ompe) = \Delta y / (c/\ompe) = 0.1$, where $\ompe = \sqrt{4\pi n_\m{e0} e^2 / \Gamma_0 m}$ is the electron plasma frequency. $n_\m{e0}$, $e$, $\Gamma_0 = \sqrt{(U_0/c)^2 + 1}$, and $m$ are the particle number density, the positron charge, the Lorentz factor of the shear velocity, and the particle mass.
The time step is $(\Delta t)\ompe=0.1$. 
We set the thermal velocity as $v_\m{thermal}/c = 10^{-3}$. 
Although we show the results for the discontinuous initial velocity shear in this paper, we have confirmed that our main result, shock formation, does not change even though the initial shear velocity has a finite width, $U(x)/c = 10 \tanh\{ x / ( 50c/\ompe)\}$.

\begin{figure}[t]
\centering
\includegraphics[scale=0.35]{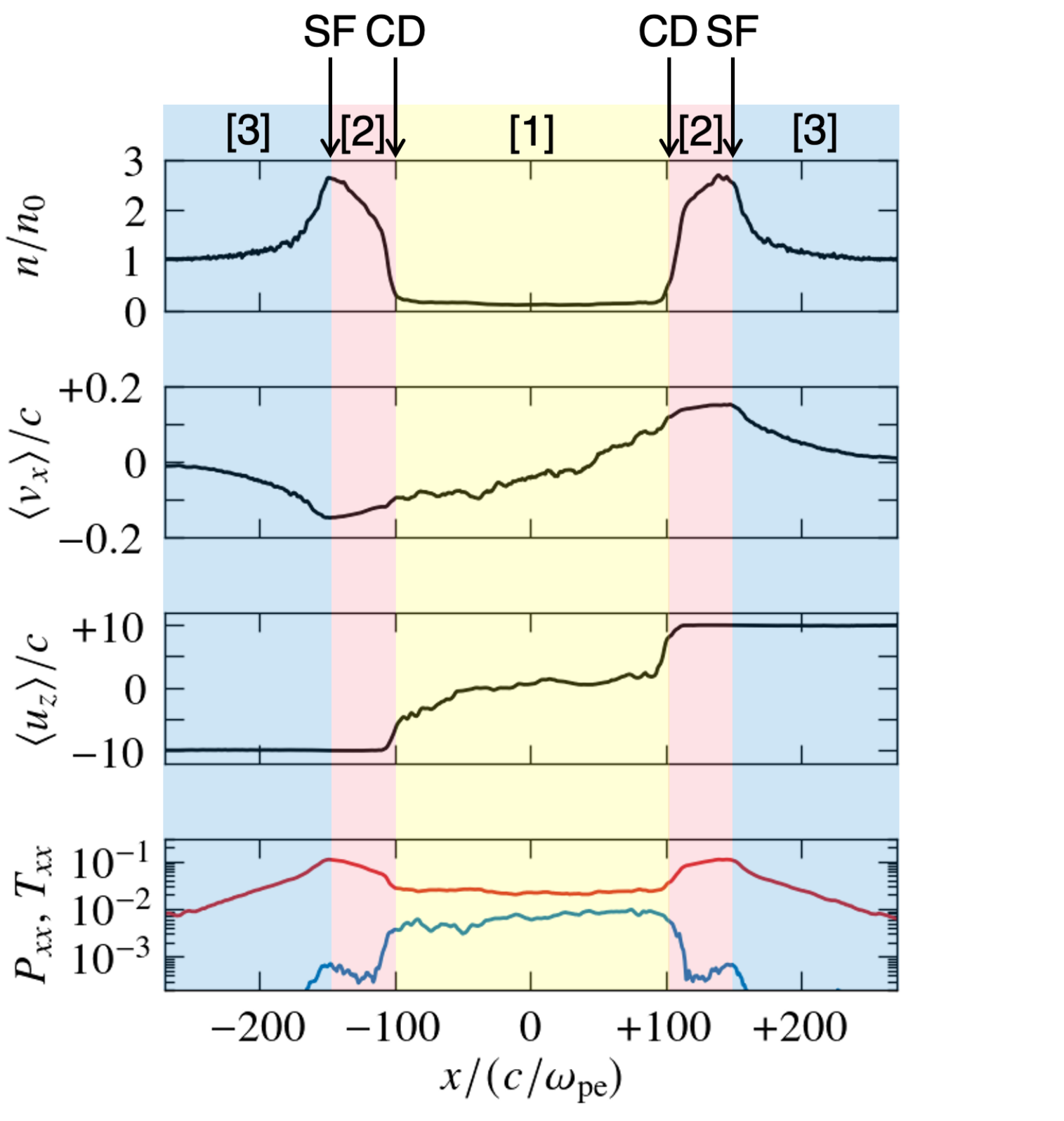}
\caption{
One-dimensional $y$-averaged profile for density (top), $x$-directional three-velocity (second from the top), $z$-directional four-velocity (third from the top), momentum tensor (bottom, red line), and Maxwell stress tensor (bottom, blue line) at $t\ompe = 600$. The regions 1, 2, and 3 indicate the low-density region, the shock downstream, and the shock upstream, respectively. CD and SF represent the contact discontinuity and the shock front, respectively.
} 
\label{fig:1D_1}
\end{figure}

\section{Result}
Fig.\ref{fig:1D_1} shows the $y$-averaged profiles of physical quantities at $t\ompe=600$, where the $x$-coordinate was adjusted so that the initial position of the shear layer was located at $x/(c/\ompe) = 0$. 
The top panel shows the total number density of electrons and positrons, $n$, normalized by the initial total number density, $n_0$. 
The second and third panels from the top show the $x$-directional three-velocity and the $z$-directional four-velocity of plasma. 
The fourth panel shows the $xx$-component of the normalized momentum tensor $P_{xx} = \int \dd^3 u \, (u_x u_x / \gamma) f(\vb{u}) / (\Gamma_0 - 1) n_0 m c^2$ in red and the $xx$-component of the normalized Maxwell stress tensor $(-B_x^2+B_y^2+B_z^2-E_x^2+E_y^2+E_z^2) / 8\pi(\Gamma_0-1)n_0mc^2$ in blue. 
Here, $f(\vb{u})$, $\vb{v}$, $\vb{u}$, $\gamma$, $\vb{B}$, and $\vb{E}$ are the distribution function, the particle three-velocity, the particle four-velocity, its Lorentz factor, the magnetic field, and the electric field, respectively. 
The symbol $\langle \, \rangle$ represents a quantity averaged in the plasma.

After the AI grows, the kinetic energy of the shear flow around the shear layer ($x/(c/\ompe)=0$) is converted to the electromagnetic and particle thermal energies, so that the pressure around $x=0$ becomes much larger than the initial value because the shear velocity is supersonic. 
As a result, plasmas on both sides of the shear layer are strongly pushed out perpendicular to the shear layer, leading to formation of collisionless shocks on both sides as shown in the top panel of Fig.\ref{fig:1D_1}. 
The speed of the shocks is $v_\m{sh}/c \sim 0.25$ in the simulation frame at $t\ompe=600$. 
Three regions with different characteristics are formed, as shown in Fig.\ref{fig:1D_1}: low-density dissipated region with $u_z/c\sim0$ (1), high-density and high-pressure shocked regions with $u_z/c=\pm10$ (2), and shock upstream regions with $u_z/c=\pm10$ (3). 
These regions are clearly separated by shock fronts (SF) and contact discontinuities (CD). 
This structure is not observed in the long-term evolution of the hydrodynamic KHI because the hydrodynamic KHI occurs only in a subsonic  velocity shear.

\begin{figure}[t]
\centering
\includegraphics[scale=0.40]{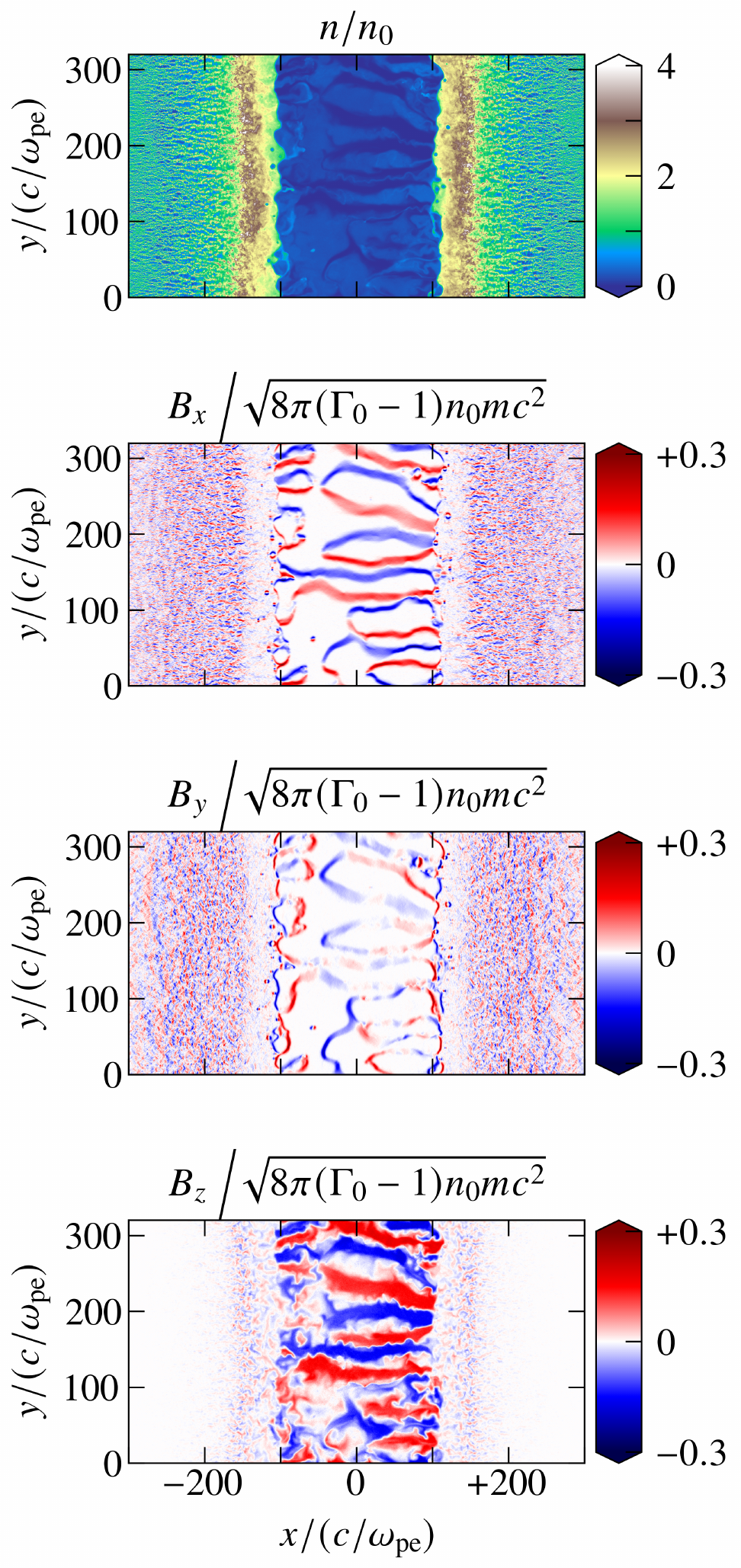}
\caption{
Two-dimensional profile at $t\ompe=600$ for the density (top) and the $x$, $y$, and $z$-components of the magnetic field (second, third, and fourth).
} 
\label{fig:2D_1}
\end{figure}

\begin{figure}[t]
\centering
\includegraphics[scale=0.35]{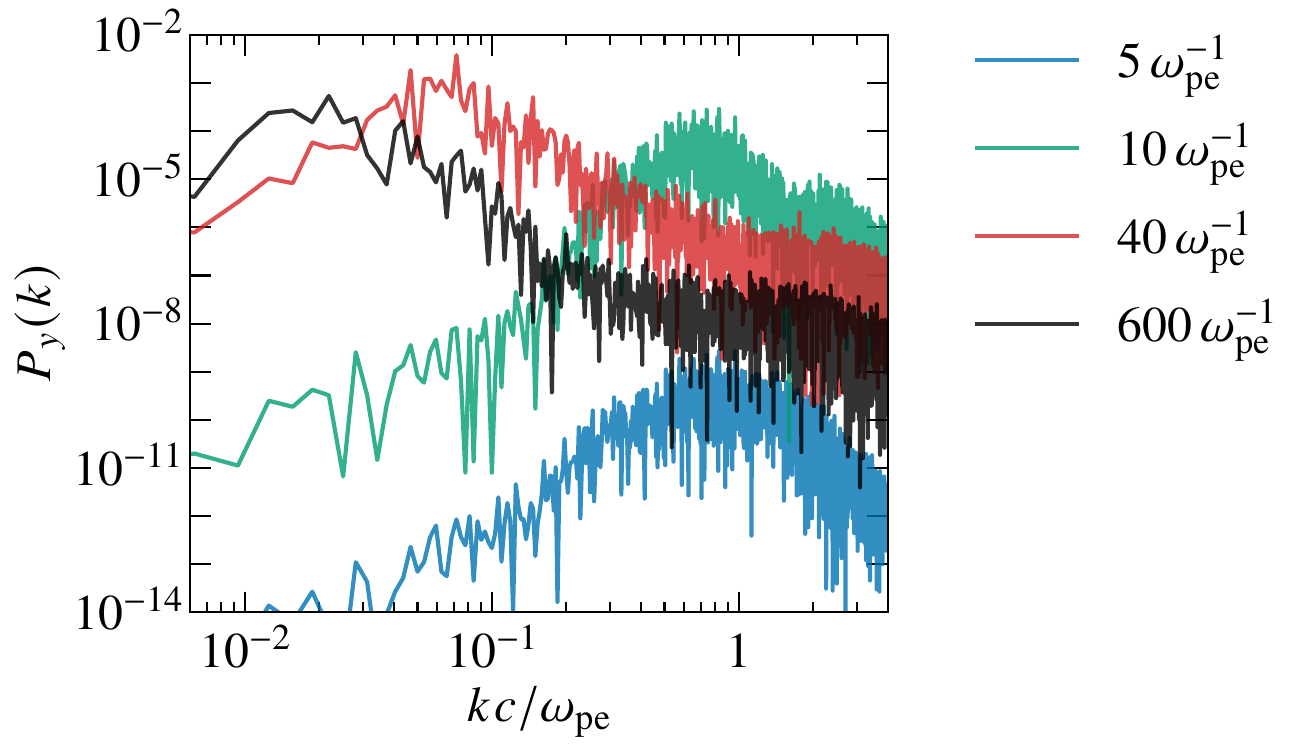}
\caption{
Power spectrum of the $x$-component of the magnetic field at $x/(c/\ompe)=0$ shown for each time.
} 
\label{fig:Spec}
\end{figure}

Fig.\ref{fig:2D_1} shows the two-dimensional profile of the density (top panel), the $x$, $y$, and $z$-components of the magnetic field (from the second panel to the fourth panel) at $t\ompe=600$. 
The density and magnetic field have small-scale fluctuations in the shock upstream region ($|x/(c/\ompe)|>150$), whereas they are larger-scale in the central low-density region.
Fig.\ref{fig:Spec} shows the $y$-directional power spectrum of $B_x$ at $x/(c/\ompe)=0$. In the early phase ($t\ompe=5-10$), the power at the inertial scale increases exponentially due to the linear phase of the AI \citep{Alves+15}. 
In the later phase ($t\ompe=40-600$), the peak wavelength becomes larger, which corresponds to the large-scale structure as shown in Fig.\ref{fig:2D_1}.
Although this nonlinear evolution of the AI has been observed in previous work \citep{Kawashima+22}, the small-scale magnetic-field fluctuation in the upstream region \textred{($|x/(c/\ompe)| > 150$)} is newly observed in this work, which provides the scattering bodies required by the shear acceleration.

\begin{figure}[t]
\centering
\includegraphics[scale=0.48]{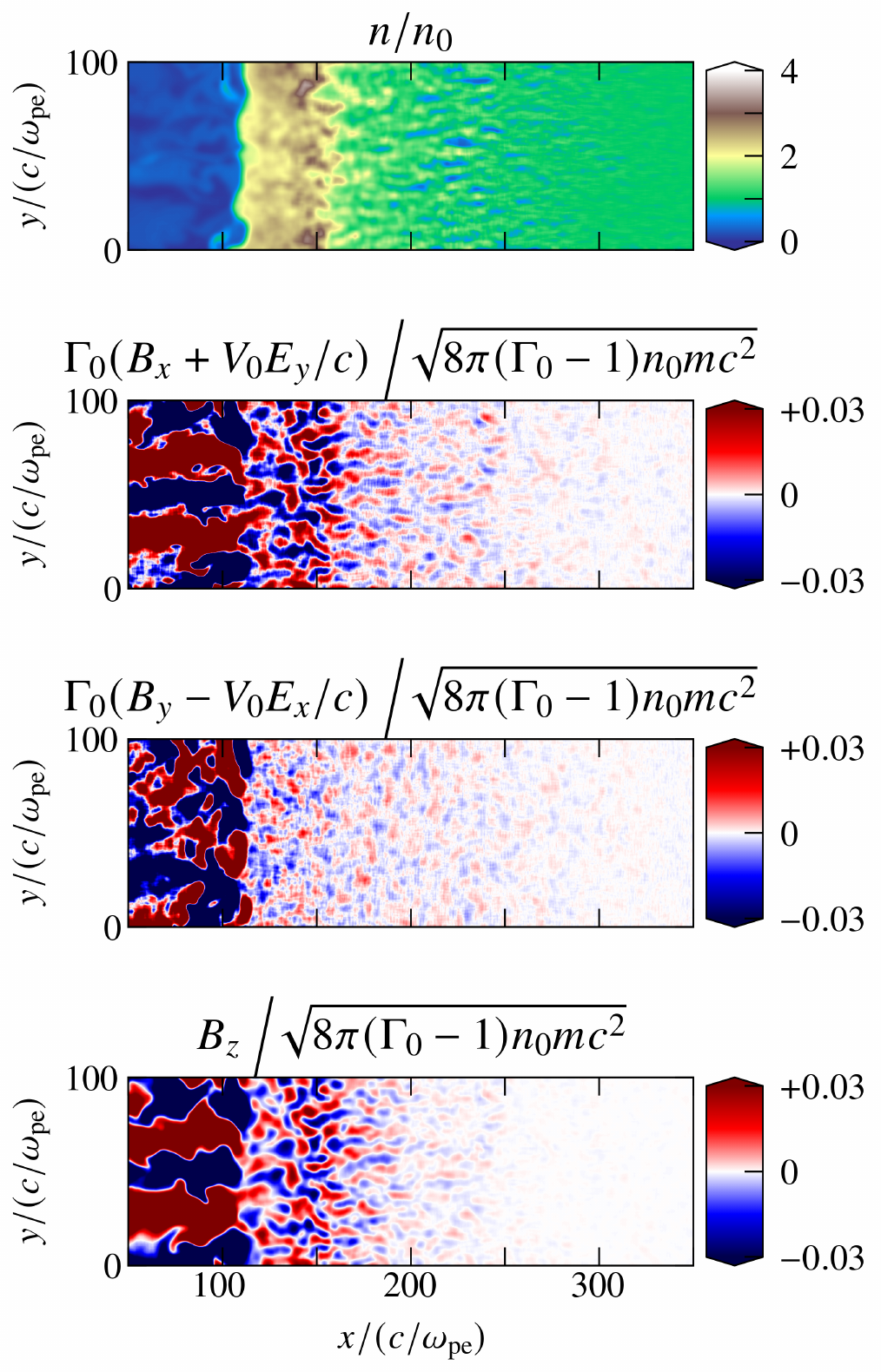}
\caption{
The same as Fig.\ref{fig:2D_1}, but zooming into $50<x/(c/\ompe)<350$ and the magnetic field is Lorentz-transformed to the upstream rest frame, where $V_0$ is the bulk velocity of the upstream plasma in the simulation frame, and $B_z^{\prime}=B_z$ is conserved.
All quantities are smoothed in $1\,c/\ompe$ and the color range is tuned to make the upstream structure clear. 
Note that the magnetic field in the low-density region is saturated. 
} 
\label{fig:2D_2}
\end{figure}

To investigate the shock structures in more detail, we plot a zoomed-in view ($50<x/(c/\ompe)<350$) in Fig.\ref{fig:2D_2}. 
The magnetic field is Lorentz-transformed to the upstream rest frame which is the frame moving with $U_z/c=+10$ in the simulation frame. 
Quantities measured in the upstream rest frame are denoted with prime. 
The Lorentz transformation of the magnetic field can be written as $B_x^\prime=\Gamma_0(B_x+V_0E_y/c)$, $B_y^\prime=\Gamma_0(B_y-V_0E_x/c)$, and $B^\prime_z=B_z$, where $V_0=U_0/\Gamma_0$ is the initial three-velocity of the shear in the simulation frame. 
The density and magnetic field structures around the shock ($150<x/(c/\ompe)<250$) are similar to those in the Weibel-mediated shock \citep{Kato07, Spitkovsky08}, \textred{except that the strength of $B_x^\prime$ is comparable to one of $B_z^\prime$}. 
In previous two-dimensional simulations of the Weibel-mediated shock \citep{Kato07, Spitkovsky08}, in which a shock wave propagates in the $x$-direction in the downstream or upstream rest frame, strong temperature anisotropy in the $x$-direction is generated in the shock transition layer, resulting in the generation of $B_z$ due to the Weibel instability in a two-dimensional system.

\begin{figure}[t]
\centering
\includegraphics[scale=0.42]{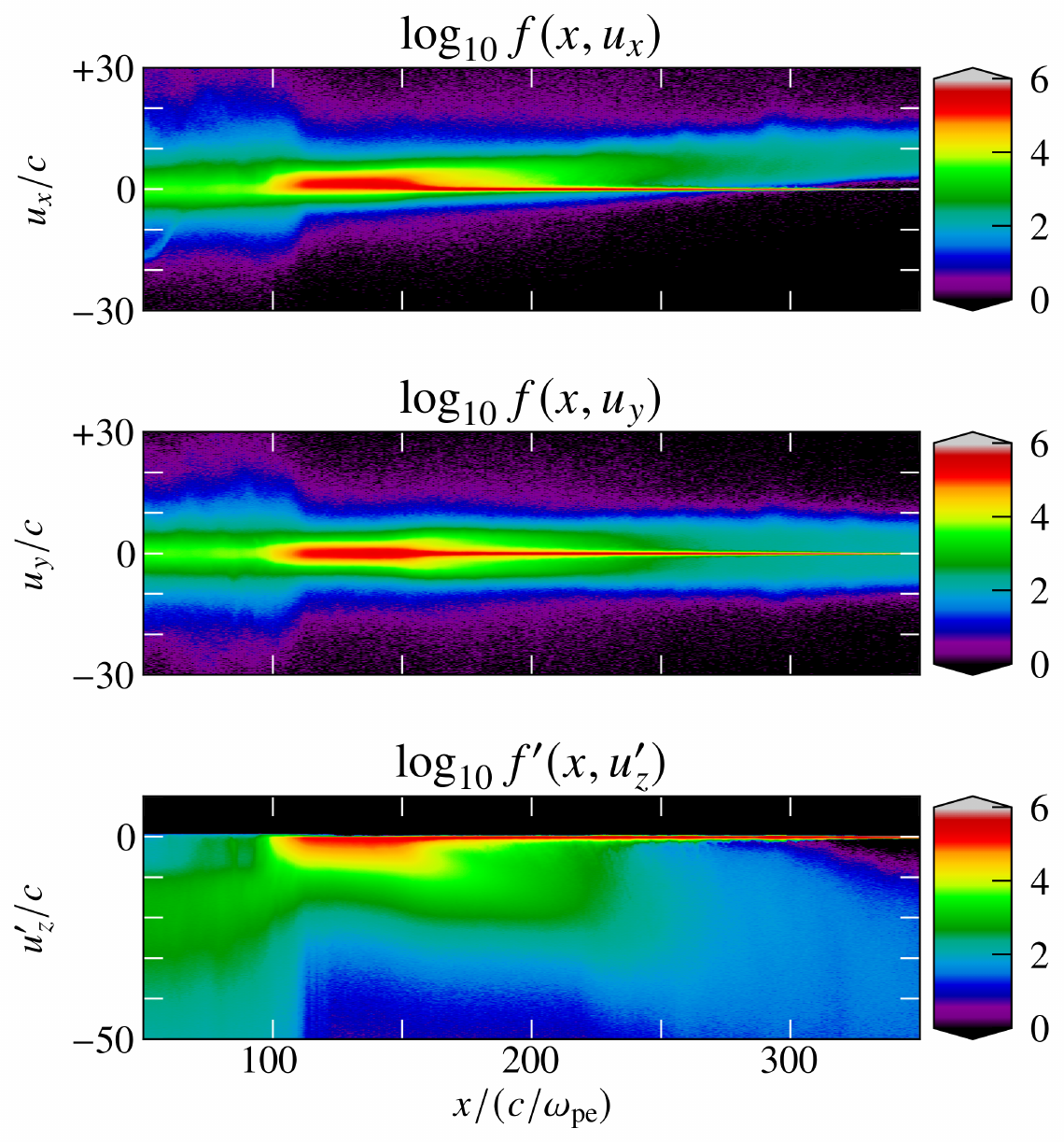}
\caption{
Phase space in the upstream rest frame at $t\ompe=600$.
} 
\label{fig:Mom}
\end{figure}

\textred{To understand the origin of the strong $B_x^\prime$ in our simulation}, we plot the phase space distribution in the upstream rest frame in Fig.\ref{fig:Mom}. 
There are particles with large negative $u_z^{\prime}$ in the shock transition region ($150<x/(c/\ompe)<250$), which was not observed in previous simulations \citep{Kato07, Spitkovsky08}. 
The origin of the large negative $u_z^{\prime}$ component can be understood by considering in which direction the shock front propagates in the upstream rest frame. 
The four-wavevector of the shock front is written as $(k_x v_\m{sh} / c, k_x, 0, 0)$ in the simulation frame. 
Performing the Lorentz transformation to the upstream rest frame, it becomes $(\Gamma_0 k_x v_\m{sh} / c, k_x, 0, -\Gamma_0 k_x V_0 v_\m{sh} /c^2)$, that is, the propagation direction of the shock wave is $(k_x^{\prime}, k_y^{\prime}, k_z^{\prime})=(k_x,0,-k_x U_0 v_\m{sh} /c^2)$ in the upstream rest frame ($|k_z^{\prime}/k_x^{\prime}| = U_0 v_\m{sh} /c^2 \approx 2.5$ for this simulation). 
Therefore, in the upstream rest frame, the shocked plasma and the plasma leaking from the downstream region have a large negative velocity in the $z^{\prime}$ direction compared with the $x^{\prime}$ direction. 
The leaking plasma with a large negative $u_z^{\prime}$ excites the Weibel instability in the shock transition region, leading to the generation of the $x^{\prime}$ and $y^\prime$ components of the magnetic field.

\section{Summary and Discussion}
In this study, we have investigated the long-term evolution of a relativistic collisionless velocity shear in an unmagnetized electron-positron plasma using a particle-in-cell simulation. 
Thanks to the long-term and large spatial-scale simulation, we found that collisionless shocks and small-scale magnetic-field turbulence are generated around the shear layer. 
A collisionless relativistic velocity shear has a large amount of free energy compared to the initial thermal energy. 
The dissipation of the free energy by the AI results in large thermal and magnetic-field pressures.
These strong pressures push out the plasma outside the shear layer, and Weibel-mediated collisionless shocks are generated.
As a result, three characteristic regions are formed as shown in Fig.\ref{fig:1D_1}: (1) a low-density region with large-scale magnetic field and $u_z\sim0$, (2) high-density regions with small-scale magnetic field and $u_z=\pm{U_0}$, and (3) initial upstream regions with small-scale or no magnetic field and $u_z=\pm{U_0}$.

It has been demonstrated that particles are accelerated in the Weibel-mediated shock \citep{Spitkovsky08,Sironi+13}, although it was not observed in this simulation owing to the limitation of the short simulation time. 
Therefore, the high-energy particles accelerated by the Weibel-mediated shock could be injected into the shear acceleration, which could be one of the solutions for the injection problem. 
In addition to the AI, the Weibel-mediated shock generates magnetic field turbulence around the shear layer, which is also required to accelerate particles by the shear acceleration. 
As shown in the third panel from the top of Fig.\ref{fig:1D_1}, our simulation showed that sharp velocity shears are still present at contact discontinuities, while a smooth shear structure is found in the low-density region.  
This is also an unexpected structure and would affect the energy spectrum of particles accelerated by the shear acceleration. 
In this work, we considered an electron-positron unmagnetized plasma with relativistic velocity shear. 
How the composition of the plasma, magnetization, and velocity of shear affect the long-term evolution of collisionless velocity shear and particle acceleration around the shear layer are also interesting questions. 
Therefore, large-scale kinetic simulations for the velocity shear will open a new window to understanding fundamental plasma physics, astrophysical plasmas, and the origin of cosmic rays.

\textred{
Finally, we discuss the eventual evolution of the shock that we found in this work. 
In the absence of subsequent dissipation of the shear flow following the initial dissipation caused by the AI, the shock propagation slows down over time.
Conversely, if subsequent dissipation of the shear flow occurs, the newly generated high-pressure plasma continues to drive the propagation of the shock. 
The long-time evolution of the shock propagation depends on how efficiently the shear flow dissipates over the long timescale.
Our simulation showed that sharp velocity shear layers have been maintained at the contact discontinuity at $t\ompe=600$ (third panel of Fig.\ref{fig:1D_1}). 
This structure is expected to be unstable, so that some additional dissipation would occurs. Further simulations and linear analyses are required to understand the long-term evolution of the contact discontinuity with the velocity shear, which will be addressed in the future.
}

\begin{acknowledgments}
We thank T. Amano for fruitful discussions. Simulations were performed on Cray XD2000 at the Center for Computational Astrophysics, National Astronomical Observatory of Japan. K.K. is supported by JSPS KAKENHI grants No. JP26KJ0915 and International Graduate Program for Excellence in Earth-Space Science (IGPEES). Y.O. is supported by JSPS KAKENHI grants Nos.JP25K00999 and JP26H00825.
\end{acknowledgments}

\end{document}